\documentclass[11pt]{article}
\usepackage{mathtools}

\usepackage[iso-8859-7]{inputenc}
\usepackage{epsfig}
\usepackage{graphicx}
\usepackage{epstopdf}
\usepackage{amsfonts}
\usepackage{amssymb}
\usepackage{amsthm}
\usepackage{amsmath}
\usepackage{multirow}
\usepackage{appendix}
\usepackage{color}
\usepackage[dvipsnames]{xcolor}
\usepackage{mathptmx}
\usepackage{slashed}
\usepackage{latexsym}

\newcommand{\eps}{{\epsilon}}
\newcommand{\beq}{\begin{equation}}
\newcommand{\eeq}{\end{equation}}
\newcommand{\bea}{\begin{eqnarray}}
\newcommand{\eea}{\end{eqnarray}}

\newcommand{\gsim}{\lower.7ex\hbox{$\;\stackrel{\textstyle>}{\sim}\;$}}
\newcommand{\lsim}{\lower.7ex\hbox{$\;\stackrel{\textstyle<}{\sim}\;$}}
\newcommand{\be}{\begin{equation}}
\newcommand{\ee}{\end{equation}}
\newcommand{\ba}{\begin{eqnarray}}
\newcommand{\ea}{\end{eqnarray}}

\usepackage{cite}
\usepackage{float}
\usepackage{amsmath} 
\usepackage{amssymb}   
\usepackage{amsthm} 
\usepackage{geometry}
\usepackage{graphicx} 
\usepackage{multirow} 
\usepackage{hyperref}
\hypersetup{
    colorlinks=true,
    linkcolor=blue!70,
    citecolor=RedViolet,
    filecolor=magenta,      
    urlcolor=RedViolet,
}

\usepackage{color}
\usepackage{float}
\usepackage{multirow}

\usepackage{lmodern}

\numberwithin{equation}{section} 

\begin{document} 
	\thispagestyle{empty}
	\begin{titlepage}
		\vspace*{0.7cm}
		\begin{center}
			{\Large {\bf  Low Scale String Theory Benchmarks for Hidden Photon Dark Matter Interpretations of the XENON1T Anomaly
			}}
			\\[12mm]
			Athanasios Karozas$^{a}$~\footnote{E-mail: \texttt{akarozas@uoi.gr}; https://orcid.org/0000-0002-1453-1086},
			Stephen F. King$^{b}$~\footnote{E-mail: {\tt king@soton.ac.uk}; https://orcid.org/0000-0002-4351-7507},
			George K. Leontaris$^{a}$~\footnote{E-mail: \texttt{leonta@uoi.gr}; https://orcid.org/0000-0002-0653-5271},
			Dimitrios K. Papoulias$^{a}$~\footnote{E-mail: \texttt{d.papoulias@uoi.gr}; https://orcid.org/0000-0003-0453-8492}
		\end{center}
		\vspace*{0.50cm}
		\centerline{$^{a}$~\it
			Physics Department, University of Ioannina,}
		\centerline{\it 45110, Ioannina, 	Greece}
		\vspace*{0.50cm}
			\centerline{$^{b}$~\it
			School of Physics and Astronomy, University of Southampton, }
		\centerline{\it SO17 1BJ, Southampton, U.K.}
		\vspace*{1.20cm}		
\begin{abstract}
		An excess of low-energy 
		electronic recoil events over known backgrounds was recently observed in the XENON1T detector,
		where $285$ events are observed compared to an expected $232 \pm 15$ events from the background-only fit to the data
		in the energy range 1-7 keV. This could be due to the beta decay of an unexpected tritium component, or possibly to new physics. One plausible new physics explanation for the excess is absorption of hidden photon dark matter relics with mass around $2.8$ keV and kinetic mixing of about $10^{-15}$, which can also explain cooling excesses in  horizontal-branch (HB) stars. Such small gauge boson masses and couplings can naturally arise from type-IIB low scale string theory. We provide a fit of the XENON1T excess in terms of a minimal low scale type-IIB string theory parameter space and present some benchmark points which provide a good fit to the data.		It is also demonstrated how the required transformation
		properties of the massless spectrum are obtained in intersecting D-brane  models. 
		\end{abstract}
		
	\end{titlepage}

\vfill
\newpage
\setcounter{footnote}{0}
{
  \hypersetup{linkcolor=black}
}
\setcounter{footnote}{0}

\section{Introduction}

The XENON Collaboration has reported an excess of low-energy 
		electronic recoil events over known backgrounds in the XENON1T detector,
		where $285$ events are observed compared to an expected $232 \pm 15$ events from the background-only fit to the data
		in the energy range $1-7$ keV~\cite{Aprile:2020tmw}. This could be due to the beta decay of an unexpected tritium component, or possibly to new physics. 
		The latter possibility has led to a plethora of theory papers~\cite{AristizabalSierra:2020edu,Boehm:2020ltd,DiLuzio:2020jjp,  McKeen:2020vpf, Bally:2020yid, Khan:2020vaf,Miranda:2020kwy,Alonso-Alvarez:2020cdv,Smirnov:2020zwf,Nakayama:2020ikz,Jho:2020sku,Bramante:2020zos,An:2020bxd,Zu:2020idx,Gao:2020wer,Lindner:2020kko,DelleRose:2020pbh,Ge:2020jfn,Takahashi:2020bpq,Okada:2020evk, Athron:2020maw, Borah:2020jzi, Choi:2020udy, Choi:2020kch, Farzan:2020dds,Arcadi:2020zni}
		which attempt to explain the XENON1T excess as due to a new particle. Most of the explanations involve the absorption of a new light weakly interacting bosonic dark matter (DM) relic, for example axionlike particles (ALPs) with spin zero, new gauge bosons $Z'$, which couple directly (but weakly) to fermions, 
				or hidden photons (also known as dark photons or paraphotons) which interact only via gauge kinetic mixing.  However there is a general issue among all the DM models which have been proposed to explain the XENON1T excess, namely how to achieve the observed DM relic density of $\Omega_{DM} h^2=0.12$,
				which can be addressed in various ways. We shall not discuss that issue further here, but just assume that a suitable (possibly multicomponent) relic density can be achieved by one of the proposed mechanisms discussed in the literature~\cite{AristizabalSierra:2020edu,Boehm:2020ltd,DiLuzio:2020jjp,  McKeen:2020vpf, Bally:2020yid, Khan:2020vaf,Miranda:2020kwy,Alonso-Alvarez:2020cdv,Smirnov:2020zwf,Nakayama:2020ikz,Jho:2020sku,Bramante:2020zos,An:2020bxd,Zu:2020idx,Gao:2020wer,Lindner:2020kko,DelleRose:2020pbh,Ge:2020jfn,Takahashi:2020bpq,Okada:2020evk, Athron:2020maw, Borah:2020jzi, Choi:2020udy, Choi:2020kch, Farzan:2020dds,Arcadi:2020zni}.
				
Turning to the different scenarios, 
there are reasons why the hidden photon explanation may be favored over either the ALP or the $Z'$ models.
To begin with, ALPs with a sufficient electron coupling, $g_{ae}=(5-7)\times 10^{-14}$, to explain the XENON1T signal, need to have extremely suppressed couplings to photons to accommodate constraints from x-ray searches~\cite{Irastorza:2018dyq}.
 Even for ALPs with a negligible coupling to photons, the electron coupling that fits the XENON1T result is outside of the $2\sigma$ region preferred by the stellar cooling anomalies. In the case of $Z'$, it may require a zero coupling to lepton doublets in order to suppress the coupling to neutrinos and so allow a sufficiently long-lived dark matter candidate~\cite{Okada:2020evk}. By contrast, the hidden photon, whose gauge kinetic term mixes with the hypercharge generator will automatically have couplings to neutrinos suppressed by powers of the ratio of the dark photon mass to the $Z$ mass, making the hidden photon practically stable. In addition, a hidden photon dark matter relic with the mass around $2.8$ keV and a kinetic mixing of about $10^{-15}$ can not only explain the XENON1T excess but can also explain cooling excesses in HB stars~\cite{Alonso-Alvarez:2020cdv,Giannotti:2015kwo}, while satisfying the astrophysical constraints. By contrast, ALPs are less well suited for simultaneously explaining the XENON1T excess and the stellar cooling anomaly for the best fit region~\cite{DiLuzio:2020jjp}, although the agreement is improved if the ALPs only constitute only a subdominant component of dark matter~\cite{Takahashi:2020bpq}. For all these reasons the hidden photon interpretation of the XENON1T excess seems to be very plausible.

It is well known that small gauge boson masses and kinetic mixings can naturally arise from low scale string theory \cite{Antoniadis:1990ew, Antoniadis:1998ig, Ghilencea:2002da, Abel:2008ai, Goodsell:2009xc, Anchordoqui:2011eg}, and a survey of possible string origins of such parameters consistent with the XENON1T excess has recently been made \cite{Anchordoqui:2020tlp}. 
As mentioned above, an explanation of the XENON1T excess through a hidden photon requires a mass in the scale of order a keV with corresponding gauge coupling $g_{X}\sim 10^{-15}-10^{-16}$.  In intersecting $D$-brane constructions the hidden photon mass and its corresponding gauge coupling are controlled by the dynamics of the background theory. However, the authors in \cite{Anchordoqui:2020tlp}, starting from a type-I string theory background have shown that obtaining such small values of masses and couplings for a directly coupling $Z'$
is challenging. Yet, they stressed that small kinetic mixing (also discussed in \cite{Benakli:2020vng}) may be possible
in string theory, which provides further theoretical motivation for the 
hidden photon explanation of the XENON1T excess.

In this paper we shall focus primarily on the well motivated 
hidden photon explanation of the XENON1T excess, and show how it may originate from low scale type-IIB string theory. We then provide a fit of the XENON1T excess in terms of a minimal low scale  type-IIB string theory parameter space and present some benchmark points which provide a good fit to the data. The remainder of the note is organized as follows. In Sec.~\ref{hidden} we review the idea of hidden photons.
In Sec.~\ref{strings} we discuss how type-IIB string theory can describe hidden photon with parameters in the desired range. In Sec.~\ref{A} we present specific models from intersecting D-branes backgrounds that can accommodate  weakly coupled gauge boson and 
hidden photons.
In Sec.~\ref{fit} we provide a fit of the XENON1T excess in terms of the parameters of minimal low scale type-IIB string theory 
and present some benchmark points which provide a good fit to the data.	
Finally section~\ref{conclusion} concludes the paper.	
Appendix~\ref{app} provides additional details for model building in the context of intersecting D-branes.

\section{Hidden photons}
\label{hidden}
Hidden photons (also known as dark photons or paraphotons)~\cite{Okun:1982xi,Holdom:1985ag,Foot:1991kb,Jaeckel:2013ija}
are defined to be the vector boson of an extra gauged $U(1)_X$ under which no Standard Model (SM) particle carries charge. 
The only coupling to the SM is via gauge kinetic mixing with hypercharge $U(1)_Y$~\cite{Holdom:1985ag}.
Below the electroweak symmetry breaking scale, the mixing is with the QED $U(1)_Q$ gauge kinetic term, 
\begin{equation}
\label{eq:kinetic:lagrangian}
    {\mathcal{L}}=-\frac{1}{4}(F^{\mu\nu})^2-\frac{1}{4}(X^{\mu\nu})^2-\frac{1}{2}\varepsilon F^{\mu\nu}X_{\mu\nu}-\frac{1}{2}m^{2}_{X}(X^{\mu})^2-j^{\mu}A_{\mu}~,
\end{equation}
where the photon field $A^{\mu}$ has field strength $F^{\mu\nu}$, the  
hidden photon field $X^{\mu}$ has field strength $X^{\mu\nu}$,
the explicit mass term $m_X$ for the hidden photon can emerge from a Higgs or St{\"u}ckelberg mechanism,
and $j^{\mu}$ represents interactions between the SM particles and the ordinary photon. 

After expressing the photon and dark photon fields in a canonically normalized kinetic basis, the two canonically normalized fields are no longer mass diagonal, and the mass matrix needs to be diagonalized. After diagonalizing the mass matrix one arrives at two mass eigenstates, the massless photon $\gamma_1$ and the massive dark photon $\gamma_2$, which are different from the original fields 
$A^{\mu}$ and $X^{\mu}$, the main difference being that the redefined dark photon has now very small couplings to charged particles.
The effect of all this can be thought of as a
field redefinition $A^{\mu}\to A^{\mu}-\varepsilon X^{\mu}$~\cite{Koren:2019iuv}, in which the kinetic mixing term has been traded for a direct interaction of the hidden photon with the electrically charged SM particles $j^{\mu}A_{\mu}\to j^{\mu}(A^{\mu}-\varepsilon X^{\mu})$,
so the interaction with electrons has a strength $\varepsilon e$, where $e$ is the electromagnetic charge. 
Since the hidden photon originates from a hypercharge mixing, it will also mix with the $Z$ boson of the SM.
However such mixing effects are in practice negligible, being suppressed by powers of the ratio of masses 
$m_X/M_Z$. Thus for example, the decay of the hidden photon into neutrinos, via $Z$ boson mixing, will be highly suppressed.

In order to explain the XENON1T excess we need keV mass hidden photons with very small kinetic mixings of the order of $\varepsilon \sim 10^{-15}$. There is a large literature of string theory explanations for such small masses and kinetic mixings~\cite{Dienes:1996zr,Lukas:1999nh,Abel:2003ue,Abel:2008ai,Jockers:2004yj,Blumenhagen:2005ga,Abel:2006qt,Goodsell:2009pi,Goodsell:2009xc,Goodsell:2010ie,Heckman:2010fh,Bullimore:2010aj,Cicoli:2011yh,Grimm:2011dx,Kerstan:2011dy,Camara:2011jg,Honecker:2011sm,Goodsell:2011wn,Marchesano:2014bia}. In the next section we shall develop the discussion recently provided in~\cite{Anchordoqui:2020tlp, Benakli:2020vng}, where we will see that, starting from the general considerations of a weakly coupled
(dark) gauge boson which couples directly, we are led to consider hidden photons which couple only through the 
small gauge kinetic mixing in order to account for the XENON1T excess.

 \section{Low Scale Type-IIB String Theory }
\label{strings}

	Consider a ten-dimensional type-IIB theory compactified on a six-dimensional space of volume $V_6$. The reduced Planck mass $M_{P}$, the string scale $M_{s}$, the string coupling $g_{s}$ and the internal volume $V_{6}$ are connected through the relation \cite{Benakli:2020vng}

	\be\label{eq:mplanck}
	M_{P}^{2}=\frac{V_{6}\;M_{s}^{8}}{(2\pi)^{7}\;g_{s}^2}\; .
	\ee

If the extra (dark) gauge boson resides on a $D(3+\delta)$ brane that wraps a $\delta$-cycle  with volume $V_{\delta}$, then the corresponding gauge coupling is given by \cite{Anchordoqui:2020tlp}
	
	\be \label{eq:gx1}
	g_{X}^{2}=\frac{(2\pi)^{\delta +1}\;g_{s}}{V_{\delta}\;M_{s}^{\delta}}	\; .
	\ee
	
Assume that $d$ of the six extra dimensions are large, with a common radius $R$ and the remaining (6-d) dimensions have a radius $\sim (1/M_{s})$. If also the $\delta$-cycle dimensions are a subspace of the $d$ large dimensions, then we have that:

\be 
V_{6}=(2\pi R)^{d}\;(2\pi M_{s}^{-1})^{6-d},\;\; V_{\delta}=(2\pi R)^{\delta}\; .
\ee

\noindent Replacing the volumes $V_6$ and $V_{\delta}$ above in Eqs. (\ref{eq:mplanck}) and (\ref{eq:gx1}) respectively and  combining them in order to eliminate the radius $R$ we find that:

\be \label{eq:gx2}
g_{X}^{2}=2\pi g_{s}\left(\frac{1}{2\pi g_{s}^{2}}\right)^{\frac{\delta}{d}}\left(\frac{M_{s}}{M_{P}}\right)^{\frac{2\delta}{d}}\; .
\ee

\noindent Note that, a very small gauge coupling $(g_{X}\ll 1)$ can be achieved either for a low string scale $M_{s}$ or for a small string coupling $g_{s}$. In the first case, minimal values for $g_{X}$ require maximum $\delta$. Then, for $\delta=d$  Eq. (\ref{eq:gx2}) simplifies to 

\be
g_{X}^{2}=\frac{1}{g_{s}}\left(\frac{M_{s}}{M_{P}}\right)^{2}
\label{gX}
\ee

\noindent and for $g_{s}=0.01-0.2$ with $M_{s}$ varying\footnote{Collider searches put the bound $M_{s}\gtrsim{8}$ TeV \cite{Sirunyan:2019vgj}.} from $10$ TeV to $10^{3}$ TeV we obtain $10^{-14}\lesssim{g_{X}}\lesssim{4.2\times{10^{-12}}}$.
Although weakly coupled, such a gauge coupling is not yet small enough in order to explain the observed excess in XENON1T,
which would require a gauge coupling $g_{X}\sim 10^{-15}-10^{-16}$.

\begin{figure}[t!]
\begin{center}
\includegraphics[width = 0.8 \textwidth]
{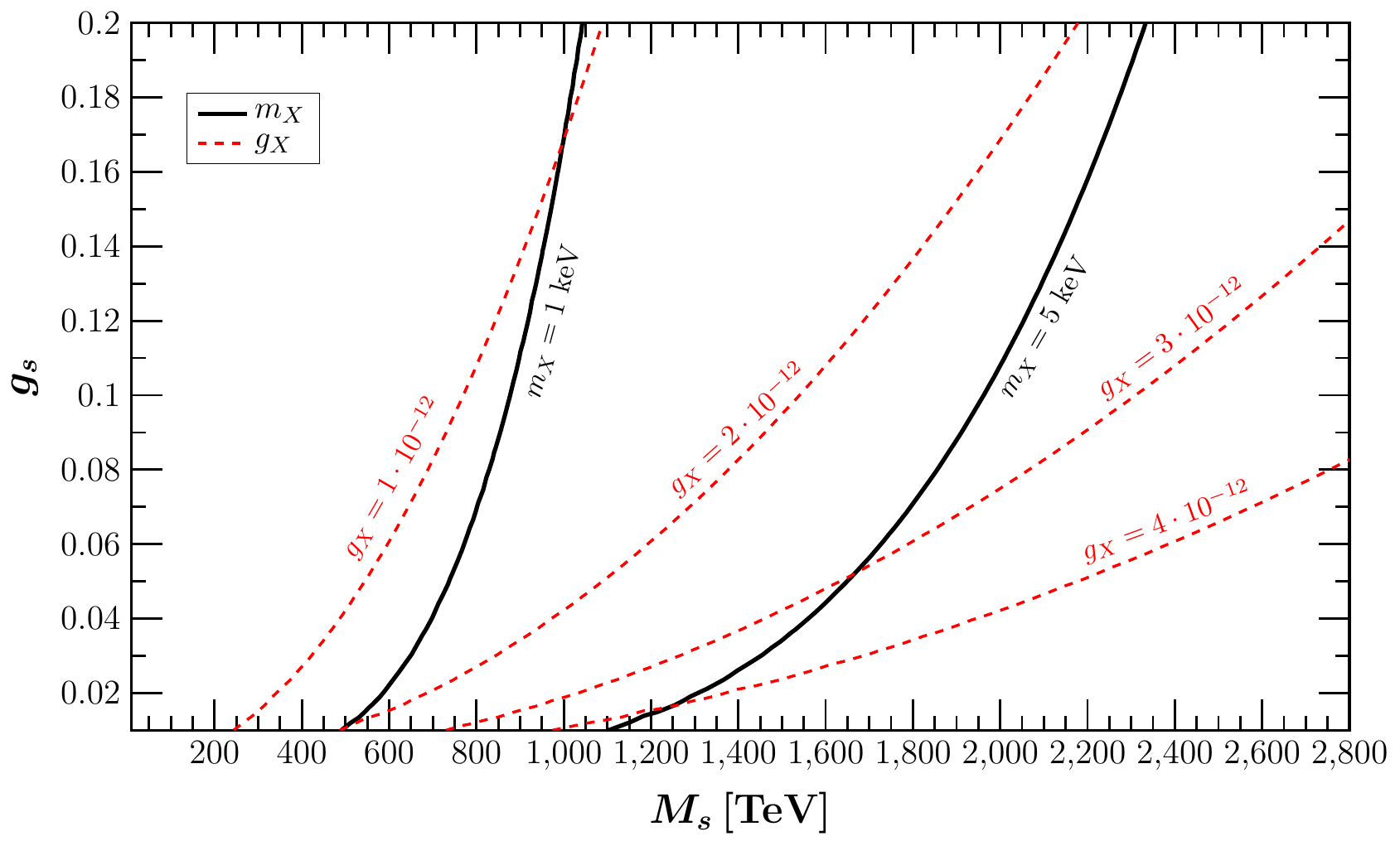}
\end{center}
\caption{\small{Contours of $m_{X}$ and $g_{X}$} in the $(M_{s},~g_{s})$ plane. For a dark gauge boson mass in the range of 1--5 keV one has $g_{X}\sim\mathcal{O}(10^{-12})$.}
\label{fig:directcouplingplot}
\end{figure}

Concerning the mass $m_{X}$ of the dark photon, this can be generated through a Higgs mechanism\footnote{An alternative scenario is that the extra $U(1)$ gauge field
becomes massive through a Stuckelberg mechanism.}. Its mass will be $m_{X}=g_{X}v_{X}$ where $v_X$ is the vacuum expectation value of the Higgs field responsible for the breaking of the extra $U(1)_X$ symmetry. Assuming $v_{X}\sim{M_{s}}$  and $\delta=d$ we obtain 

\be \label{eq:mx1}
m_{X}=\sqrt{\frac{1}{g_{s}}}\left(\frac{M^{2}_{s}}{M_{P}}\right)
\ee

\noindent which varies from 0.13 eV to 0.42 eV for $g_{s}=0.2$ to $g_{s}=0.01$ and $M_{s}=10$ TeV. Higher values for the dark photon mass can be achieved by increasing the value of $M_{s}$ which at the same time increases the value of $g_{X}$. This is shown in Fig.~\ref{fig:directcouplingplot} where we contour plot the dark photon mass $m_{X}$ and the gauge coupling $g_{s}$ in the $(M_{s},~g_{s})$ plane. As we observe, for $m_{X}$ in the range of few keV's the corresponding gauge coupling receives values, $g_{X}\sim\mathcal{O}(10^{-12})$. We see that, while $m_{X}$ has the appropriate keV value, $g_{X}$ is not small enough in order to accommodate the XENON1T results.
 Further suppression, however, might be possible through mixing effects in intersecting D-brane configurations with  additional hidden $U(1)$'s,
to be discussed in the next sections. 

Another well motivated possibility of interest to explaining the XENON1T results, 
is the hidden photon scenario.
This corresponds to the case where the extra $U(1)_X$ gauge bosons do not couple directly to the SM states 
but are only allowed to couple through the kinetic mixing with hypercharge $U(1)_Y$, as discussed in the previous section. Here we shall assume that the kinetic mixing parameter $\varepsilon$ in Eq. (\ref{eq:kinetic:lagrangian})
is generated by loops of states with masses $m_{i}$ carrying charges $q_{i},\;q^{X}_{i}$ under the two $U(1)$'s, as follows:

\be 
\varepsilon=\frac{e g_{X}}{16\pi^{2}}\sum_{i}q_{i}q^{X}_{i}\ln\frac{m^{2}_{i}}{\mu^2}\equiv \frac{e g_{X}}{16\pi^{2}}C_{\textrm{Log}}
\ee

\noindent where $\mu$ is the renormalization scale. The effective coupling to electrons discussed in the previous section 
is then identified as :

\be \label{eq:gxeff} 
g_{X,\text{eff}}=\varepsilon e =\frac{\alpha_{em}\; g_{X}}{4\pi}C_{\text{Log}} \, ,
\ee
with $\alpha_{em}=e^2/4\pi$ being the fine structure constant. Note that, the parameter $C_{\textrm{Log}}$ is model dependent, however, partial cancellations can occur leading to large suppressions in the effective coupling $g_{X,\text{eff}}$. 
\footnote{For a detailed analysis in connection with the weak gravity conjecture we refer to~\cite{Benakli:2020vng}.}

In the next section, we provide additional string theory motivation for weakly coupled gauge boson and hidden photons.
Then, using the effective gauge coupling given above and the dark photon mass given by Eq.~\eqref{eq:mx1} we interpret the XENON1T excess in the parameter space of $M_{s}$, $g_{s}$ and $C_{\text{Log}}$.

\section{D-brane configurations} \label{A}
String model building offers a variety of solutions.
Among the most popular  constructions are SM  extensions from 
intersecting D-branes~\cite{Ibanez:2001nd,Anastasopoulos:2006da}
and F-theory~\cite{Beasley:2008kw} motivated models~\cite{Romao:2017qnu,Karozas:2020zvv}, which is the 12-dimensional geometric manifestation of type-IIB string theory.

	\begin{figure}[htb!]
	\centering
	\includegraphics[scale=0.30]{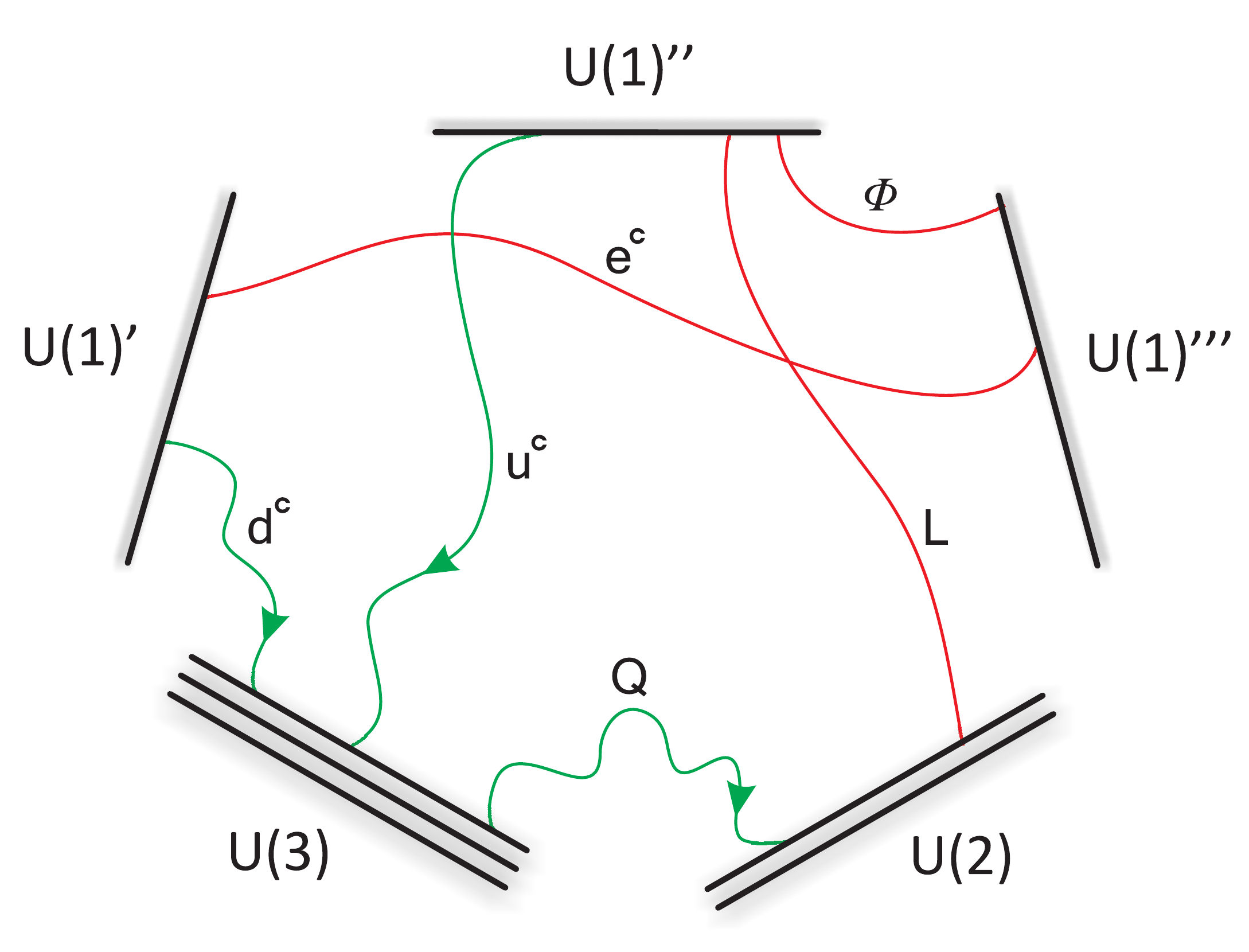}\;\;\;\;
		\includegraphics[scale=0.30]{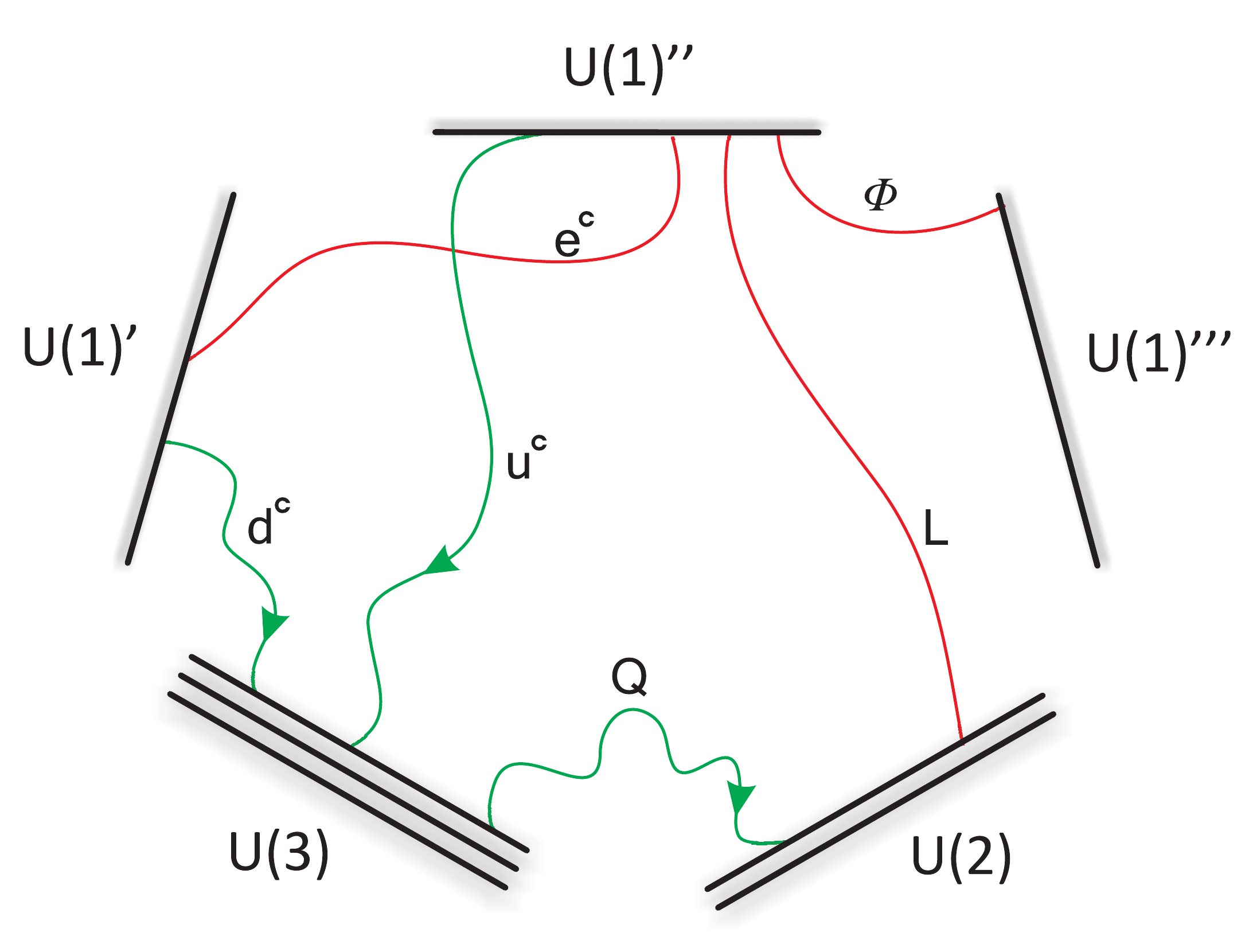}
	\caption{\small{Two possible configurations of open strings with 
		their two endpoints attached on the intersecting D-branes. Higgs fields -not shown- are represented with strings with one 
		endpoint attached to $U(2)$ stack and the other one to any of the three $U(1)$ D-branes.
		The $U(1)'''$ brane is associated with the very weak gauge coupling $g_X$ in Eq.\ref{gX}. 
		The left panel describes the model in~\cite{Okada:2020evk} where only $e^c$ couples to $U(1)_X$.
		The right panel corresponds to the case of the hidden photon, where none of the SM particles couples to $U(1)_X$.}} 
	\label{fig1:Dbranemodels}
\end{figure}

 Here  is a brief description how   the gauge  symmetry and the spectrum arise in intersecting D-brane models in type-I string theory~\footnote{The analysis in Sec.~\ref{strings} can be converted from type-II to type-I by multiplying the right-hand side of Eq.~(\ref{eq:mplanck}) with a factor of 2.}.
 The fundamental object  is the brane stack, consisting of a certain number of parallel, almost coincident D-branes.  
 A single D-brane carries a $U(1)$ gauge symmetry  whereas a stack of $n$ parallel branes gives rise to a
$U(n)\sim SU(n)\times U(1)$ gauge group.  When stacks  intersect each other chiral fermions are represented by strings jostling in the 
intersections sitting in singular points of the transverse space. For definiteness, the compact space is supposed 
to be a six-dimensional torus $T^6=T^2\times T^2\times T^2$. Chirality arises when the brane-stacks  are wrapped on a 
torus~\cite{Blumenhagen:2000wh} and the multiplicity  of fermion generations is topological invariant 
depending on the homology classes and is  given by a formula involving 
the two distinct numbers of brane wrappings around the two circles of the torus. Thus, for two stacks $n, m$, the gauge 
group is $U(n)\times U(m)$ while the fermions are in the bifundamental representations $(n,\bar m)_{q_n, q_{\bar m}}$, or $(\bar n,
m)_{q_{\bar n},q_m}$ where the indices refer to the corresponding $U(1)$ charges.  Obviously, if an open string is attached  on an $SU(n)$ stack  with the other end on a $U(1)_X$ brane, then the 
corresponding state  is  $(n, 1)_{(q_n,q_X)}$, where the indices  refer to the ``charges" of the state under the
Abelian symmetries  $U_n(1)\in U(n)$ and  $U(1)_X$. The minimum 
number of brane stacks required to accommodate the SM are a stack of three parallel branes implying $U(3)\to SU(3)\times U(1)$,  
another one of two parallel branes $U(2)\to SU(2)\times U(1)$, and a $U(1)$. The hypercharge is a linear combination of 
these three abelian symmetries. However, this minimal structure  cannot accommodate a dark photon with a light mass and tiny 
couplings  to ordinary matter. Evidently, a configuration of more than one $U(1)$ brane should be considered~\cite{Anchordoqui:2020tlp}. 
 
In~\cite{Gioutsos:2005uw} a viable intersecting  D-brane set up 
has been considered which is capable of interpreting the XENONT1 effects.
It consists of $U(3)$ and $U(2)$ stacks giving rise to the 
non-Abelian factors of SM, and three $U(1)$ branes. In this case, the symmetry of the emerging model is
\be 
SU(3)_c\times U(1)_c\times SU(2)_L\times U(1)_L\times U(1)^{3}\; .
\label{SM3U1}
\ee 

Two possible configurations of open strings with 
their two endpoints attached on the intersecting D-branes and emerging symmetry given by (\ref{SM3U1}) graphically illustrated in Fig.~\ref{fig1:Dbranemodels}. Higgs fields--not shown--are represented with strings with one 
endpoint attached to $U(2)$ stack and the other one to any of the
three $U(1)$ D-branes.

Next, the salient features are described: depending on 
the specific spectrum of the model, several $U(1)$'s could be anomalous and a generalized Green-Schwarz mechanism is implemented to 
cancel the anomalies of the corresponding Abelian factors. 
The hypercharge generator of the model  is 
a linear combination of $U(1)$ factors  contained in the
gauge symmetry~(\ref{SM3U1}). 
In principle all $U(1)$'s could take part in this combination,
hence the hypercharge is written as a linear combination 
\be 
Y= k_3 Q_3+k_2 Q_2+ \sum_{j=1}^3 c_j Q_{1,j}
\label{Hyperc}
\ee 

\noindent where $k_{3}$, $k_{2}$ and $c_{j}$ are real constants.

Each state is represented  by an open string with its two ends attached to two brane stacks and
as a result, they carry two $U(1)$ charges. 
Assigning the appropriate charges should take into account  the 
  orientation of the strings, i.e., where the open strings begin
and end. 

All possible hypercharge embeddings of the spectrum can be found as follows. In the
example shown in Table \ref{tab1:charges}, the  SM states are characterized by 
the charges under the abelian factors related to the  symmetries
of each stack. Thus, the quark doublet is represented by a string 
with its endpoints on $U(3)_c$ and $U(2)_L$ stacks, hence the 
charges are shown in the corresponding entries of the table. 
The coefficients  $\eps_k$  take the values $\pm 1$ depending on whether the string ends
on the brane or its mirror (under orientifold  action).
\begin{table}
	\centerline{   
		\begin{tabular}{c|cccccccccc}
			\hline\hline
			\multirow{2}{*}{Spectrum} &
			\multicolumn{4}{c}{Charges under the abelian symmetries}&
			\\
			& $U(1)_C$ &$ U(1)_L$ & $ U(1)'$ &$ U(1)''$ & $ U(1)'''$  \\
			\hline
			$Q$ & $+1$& $\eps_1$  & $0$  & $0$& $0$   \\
			$u^c$ & $-1$& $0$  & $0$  & $\eps_2$& $0$    \\
			$d^c$ & $-1$& $0$  & $\eps_3$   & $0$ &  $0$ \\
			$L$ & $ \phantom{+}0$& $\eps_4$   & $0$ & $\eps_5$  & $0$   \\
			$	e^c$ & $\phantom{+}0$& $0$  & $\eps_6$  & $0$  & $\eps_7$ \\
			$ \Phi$ & $\phantom{+}0$& $0$   & $0$  & $\eps_8$ & $\eps_9$ \\
			\hline\hline
	\end{tabular} 	}
	
	\caption{\small{The ``charge" assignments of the spectrum originating from the D-brane configuration of Fig. \ref{fig1:Dbranemodels} (left panel).	}}\label{tab1:charges}
\end{table}
According to  the previous discussion, in order to accommodate a light neutral  boson $Z'$, 
its associated left-handed neutral currents involving quark and lepton doublet fields should
non exist. This is possible if the light boson is associated with the $U(1)'''$ brane 
of the left panel in Fig.~\ref{fig1:Dbranemodels}, where only $e^c$ couples. 
Using (\ref{Hyperc}), the following equations determine the hypercharges of the various SM
states
\ba 
\begin{aligned}
Q^{\phantom{c}}:\quad &\phantom{+} k_3+\eps_1 k_2 &=&\phantom{+}\frac 16
\\
u^c:\quad&-k_3+\eps_2 c_2&=&-\frac 23
\\
d^c:\quad&-k_3+\eps_3 c_1&=&\phantom{+} \frac 13
\\
L^{\phantom{c}}:\quad&\;\;\eps_4k_2+\eps_5 c_2&=&-\frac 12
\\
e^c:\quad&\;\;\eps_6c_1+\eps_7 c_3&=&\phantom{+}1
\\
\Phi :\quad&\;\;\eps_8c_2+\eps_9 c_3&=&\phantom{+}0\; .
\\
\end{aligned}
\ea 
The last entry corresponds to a SM neutral singlet, $\Phi$. This equation can be solved trivially by assuming that $c_{2}=c_{3}=0$ (or $c_{2}=c_{3}$ and $\eps_{8}=-\eps_{9}$).

Among the various  solutions of the above system,  a promising one  is:

\begin{equation}
\begin{split}
k_{3}&=\frac 23,\; k_2=\frac 12,\; c_1=1,\; c_2=0,\;c_3=0,\\
\eps_{1}&=-1,\;\eps_4=-1,\; \eps_3=+1,\;\eps_6=+1
\end{split}
\end{equation}

\noindent where the $\eps_{i}$  (not shown) can receive any of the values $\pm{1}$. Note that $c_{2}=c_{3}=0$ which means that $U(1)''$ and $U(1)'''$ decoupled from the definition of the hypercharge. All the possible solutions with $c_{2}=c_{3}=0$ are listed in Appendix \ref{app}.

We turn now into the Higgs fields  and the Yukawa sector of the model. The requirement for tree-level up and down quark terms  fixes the $U(1)$ charges of the MSSM Higgs doublets $H_{u}$ and $H_{d}$ respectively. Their charges are presented in Table \ref{tab:Higgscharges}. Notice that a charged lepton renormalizable operator of the form $L\;e^{c}\;H_{d}$, is not invariant under the various $U(1)$ symmetries of the model. For $\eps_{8}=-\eps_{5}$ and $\eps_{9}=-\eps_{7}$, the charged lepton masses appear through nonrenormalizable operators of the form  $\sim\frac{1}{M} L\;e^{c}\;H_{d}\;\Phi$.    
\begin{table}[h]
	\centerline{   
		\begin{tabular}{c|cccccccccc}
			\hline\hline
			\multirow{2}{*}{Spectrum} &
			\multicolumn{4}{c}{Charges under the abelian symmetries}&
			\\
			& $U(1)_C$ &$ U(1)_L$ & $ U(1)'$ &$ U(1)''$ & $ U(1)'''$  \\
			\hline
			$H_u$ & $0$& $1$  & $0$  & $-\eps_{2}$& $0$   \\
			$H_d$ & $0$& $1$  & $-1$  & $0$& $0$    \\
			$\Phi$ & $0$& $0$  & $0$   & $-\eps_5$ &  $-\eps_{7}$ \\
			\hline\hline
	\end{tabular} 	}
	
	\caption{\small{The ``charge'' assignments of the Higgs states originating from the D-brane configuration of Fig.~ \ref{fig1:Dbranemodels} (left panel).	}}\label{tab:Higgscharges}
\end{table}

\noindent With the charge assignments shown in Table \ref{tab:Higgscharges} we have the following superpotential terms for the quark and charged lepton sectors of the model:

\be
W\supset y_{u} Q\;u^{c}\;H_{u}+y_{d} Q\;d^{c}\;H_{d}+\frac{1}{M} L\;e^{c}\;H_{d}\;\Phi \, .
\ee
Right-handed neutrino singlets can be represented by string starting and ending on the same ($U(1)''$) brane. In this 
case,  a coupling  $L\nu^c H_u $ is possible, whereas a seesaw mechanism can take place either with higher order nonrenormalizable
terms  involving $\nu^c\nu^c$ and/or their KK excitations.

\section{Fit to the XENON1T signal}
\label{fit}

Hidden photons couple with electrons in the same way as photons up to a suppression factor $g_{X,\text{eff}}$. Then, the dark photon rate at the XENON1T detector can be expressed as~\cite{An:2014twa}
\begin{equation}
N = \mathcal{E} N_{T} g_{X,\text{eff}}^2 \frac{ \rho_{\text{DM}}}{m_X m_\mathcal{N}}  \sigma_{\text{pe}} \, ,
\label{eq:signal}
\end{equation}

\noindent where  $\mathcal{E}=0.65$~ton.yr and $N_{T}$ represent the exposure and the number of atomic targets per ton at the detector, while the dark photon mass $m_X$ and the gauge coupling $g_{x,\text{eff}}$ are given in terms of the string parameters according to Eqs.(\ref{eq:mx1}) and (\ref{eq:gxeff}), respectively. Here, $\rho_\text{DM}= 0.3$~$ \mathrm{GeV cm^{-3}}$ is the DM density, $m_\mathcal{N}$ denotes the nuclear mass, while $\sigma_\text{pe}$ represents the SM photoelectric cross section evaluated at $E_\gamma = m_X$. The latter accounts for the absorption of an ordinary photon by the target
atoms. Due to the finite energy resolution of the XENON1T detector, the observed signal will be recorded as a function of the reconstructed recoil energy $T_{rec}$ through a convolution of the monochromatic rate given in Eq.(\ref{eq:signal}) with the smearing function $\mathcal{G}(T_{rec}, m_X) $ and the detector efficiency $\mathcal{A}(T_{rec})$, as
\begin{equation}
\frac{dN}{dT_{rec}} = \mathcal{A}(T_{rec})  \mathcal{G}(T_{rec}, m_X) N \, ,
\label{eq:expected-signal}
\end{equation}
\noindent where, $\mathcal{G}(T_{rec}, m_X) $ is approximated by a normalized Gaussian function with~\cite{Aprile:2020yad}:
\[ \sigma/T_{rec}   = 0.3171/\sqrt{T_{rec} \, \mathrm{[keV]}} + 0.0015\; . \]
\begin{figure}[t]
\begin{center}
\includegraphics[width = 0.48 \textwidth]{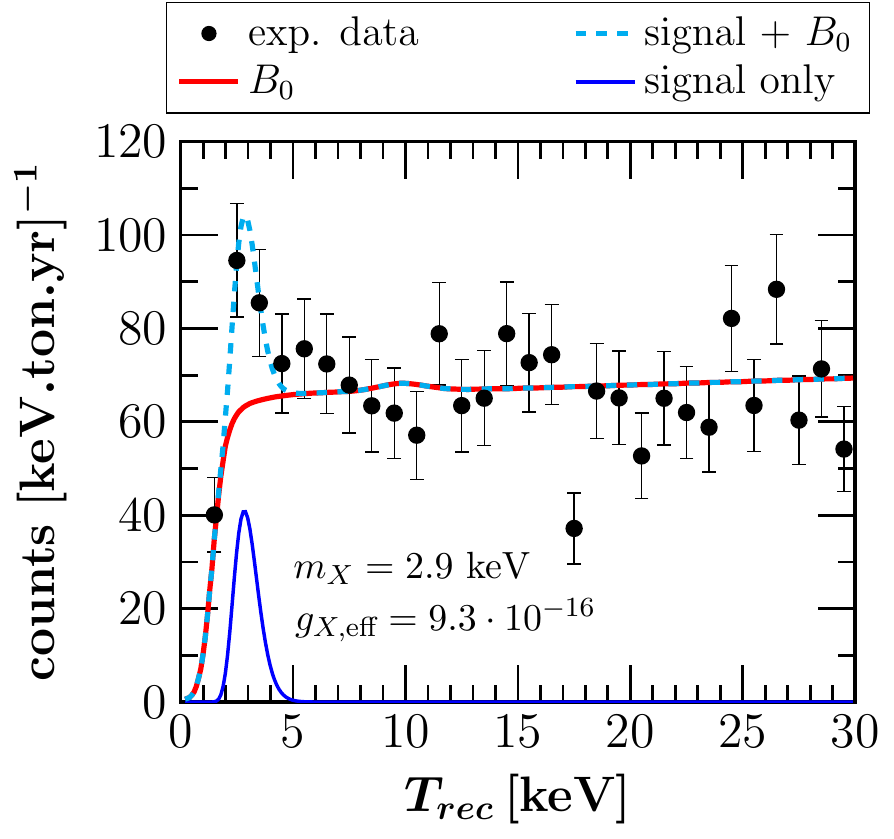}
\includegraphics[width = 0.48 \textwidth]{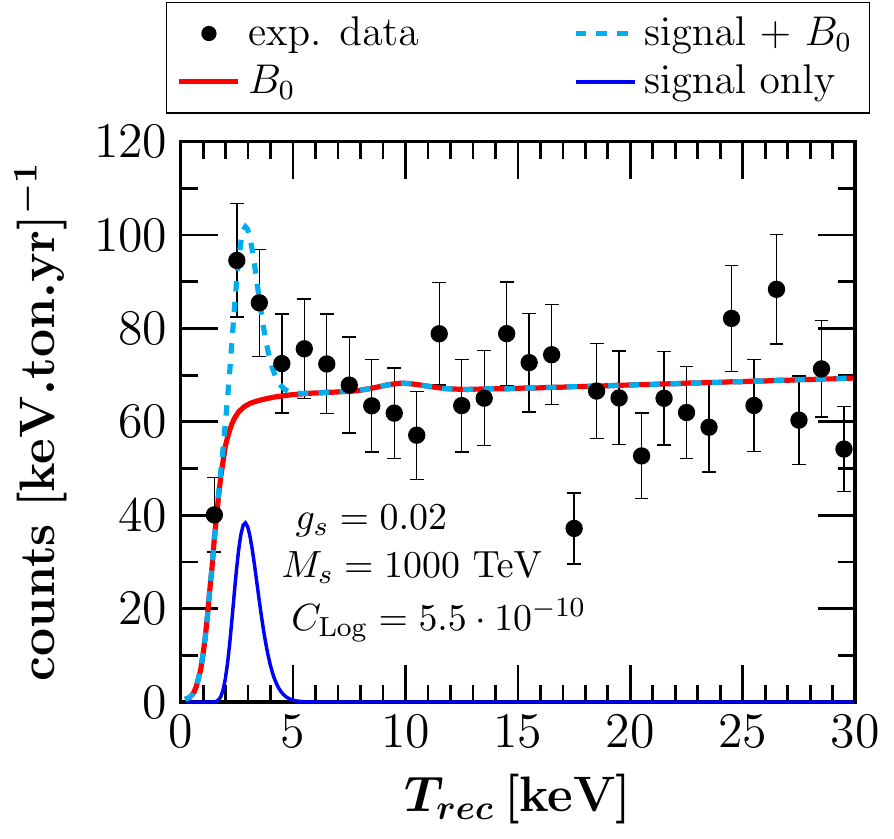}
\end{center}
\caption{\small{{\bf Left:} Expected number of events in the presence of dark photon DM scattering at XENON1T assuming the effective parameters $m_X= 2.9$~keV and $g_{X, \text{eff}}=9.3 \cdot 10^{-16}$. {\bf Right:} Same as left but assuming the string parameters $g_s = 0.02$  and $M_s =1000$~TeV and $C_\text{Log} = 5.5 \times 10^{-10}$. A comparison with the observed signal is also given.}}
\label{fig:excess}
\end{figure}

The presence of a dark photon leads to a signal enhancement at 1--4 keV recoil energy that may  explain the low energy excess events observed at XENON1T. The left panel of Fig.~\ref{fig:excess} illustrates the expected event spectrum evaluated under the assumption of effective parameters for describing the dark photon mass and kinetic mixing coupling, while the obtained results are compared to recent the XENON1T data and the background $B_0$\footnote{A discussion on possible background sources in addition to $B_0$ is given in~\cite{Bhattacherjee:2020qmv}.}. In agreement with previous estimates~\cite{Nakayama:2020ikz}, our best fit (see below) implies that a very tiny coupling $g_{X,\text{eff}}= 9.3 \cdot 10^{-16}$  and a dark photon mass of $m_X=2.9$~keV fit nicely the data. Then, we are interested to explore the prospect of interpeting the XENON1T anomaly in terms of the Low String Scale model described above. The right panel of Fig.~\ref{fig:excess} shows the corresponding expected recoil spectrum by assuming the benchmark values $g_s = 0.02$ and $M_s = 1000$~TeV and $C_\text{Log} = 5.5 \times 10^{-10}$.  Evidently, the absorption of dark photons by the atomic nuclei of the XENON1T detector is sufficient to produce a low-energy bump even for tiny couplings and hence being free of any tension with other known constraints~\cite{Alonso-Alvarez:2020cdv}.

\begin{figure}[h]
\begin{center}
\includegraphics[width = 0.75 \textwidth]
{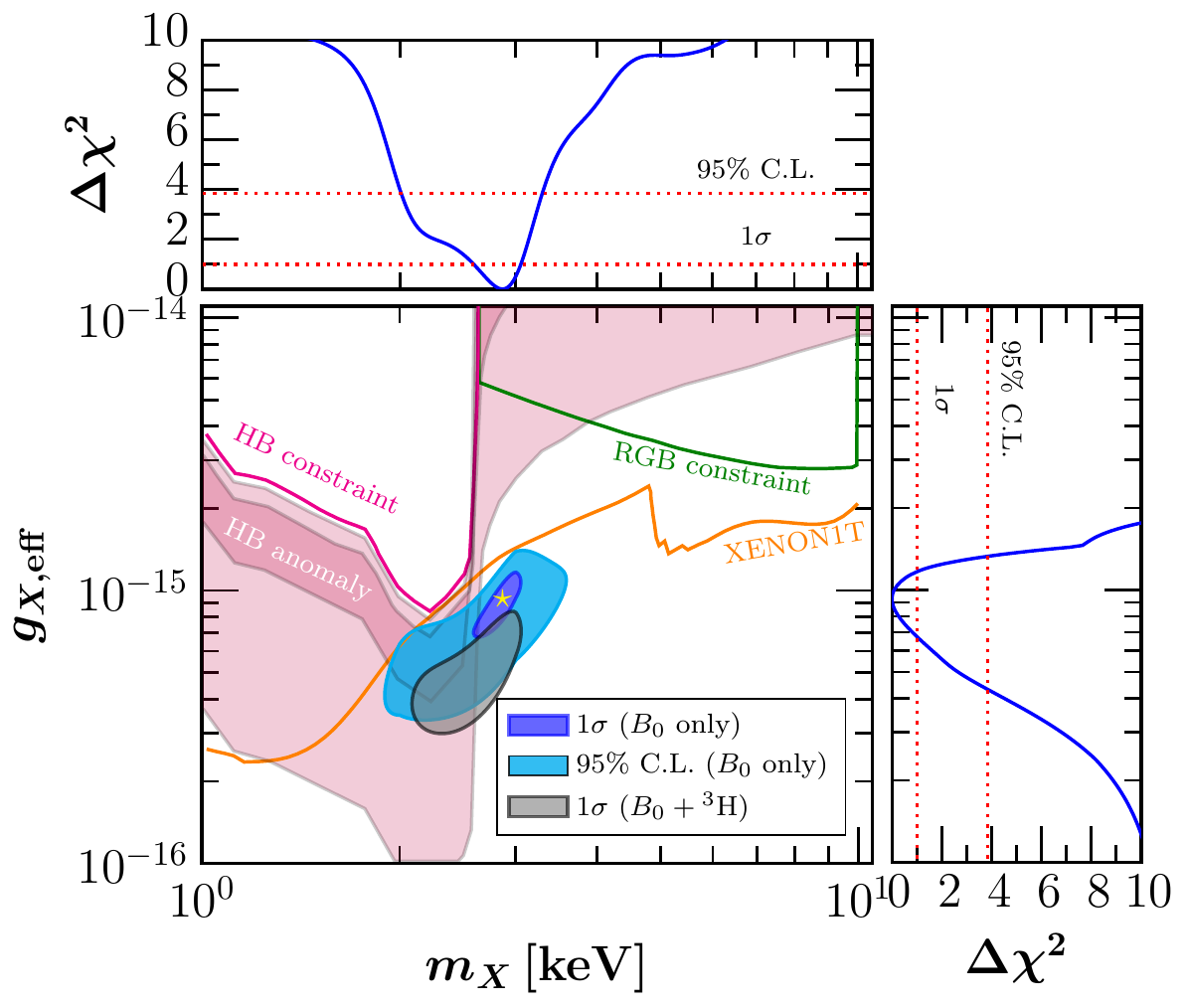}
\end{center}
\caption{\small{Sensitivity contours in the $(m_X,~ g_{X\text{eff}})$ parameter space at $1\sigma$ and $95$\% C.L. from the analysis of XENON1T data. The best fit point is also shown (astrophysical constraints are taken from~\cite{Ayala:2014pea,Giannotti:2015kwo}). $\Delta \chi^2$ profiles as a function of the effective dark photon mass and the effective coupling are also shown.}}
\label{fig:contour-eff}
\end{figure}

\begin{figure}[h!]
\begin{center}
\includegraphics[width = 0.75 \textwidth]
{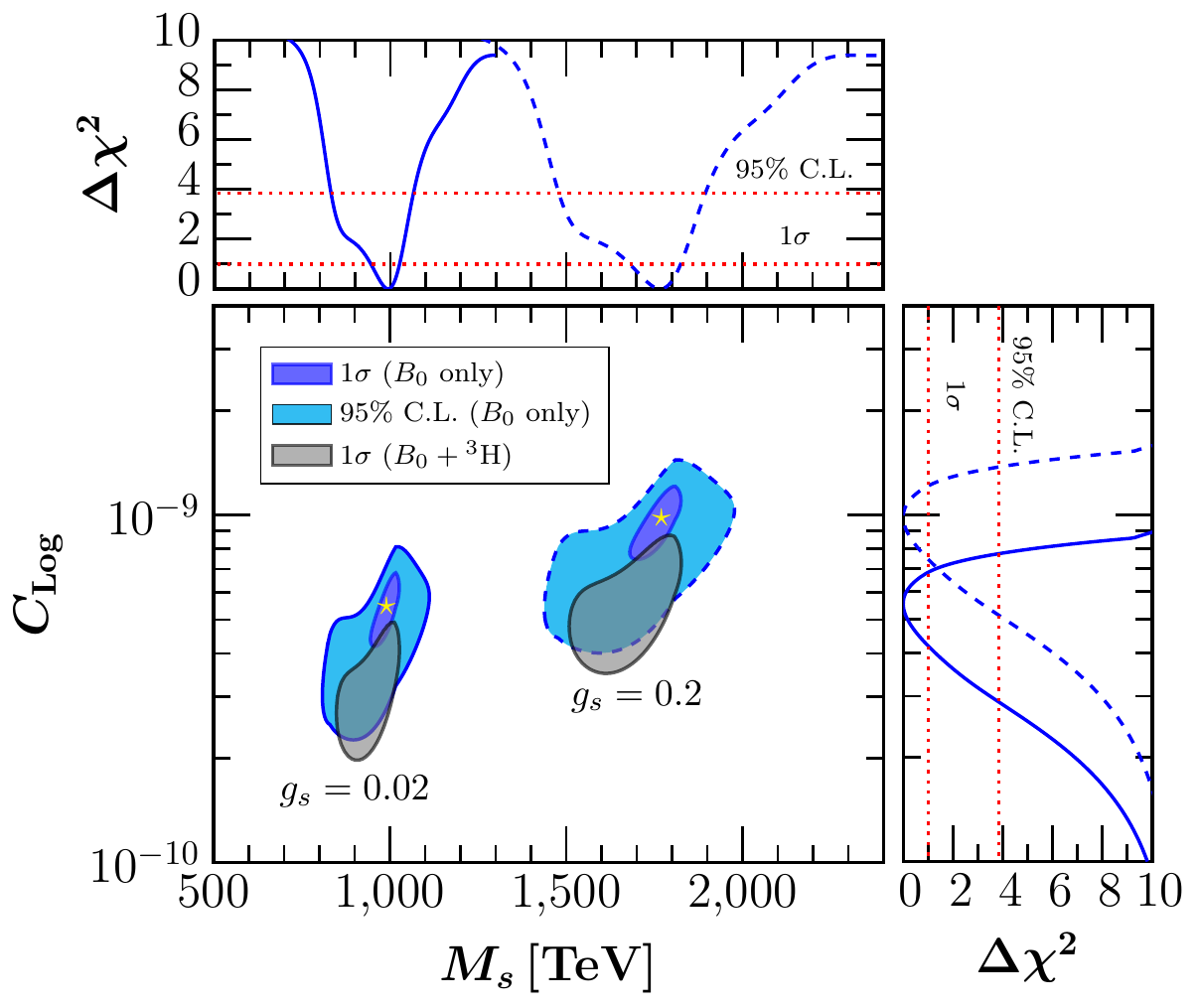}
\end{center}
\caption{\small{Same as Fig.~\ref{fig:contour-eff}, but for the string parameters $M_s$ and $C_\text{Log}$, evaluated for two benchmark values $g_s=0.02$ (solid lines) and $g_s=0.2$ (dashed lines).}}
\label{fig:contour-string}
\end{figure}

We are therefore motivated to perform a spectral fit based on the $\chi^2$ function~\cite{AristizabalSierra:2020edu}
\begin{equation}
\chi^2(g_s, M_s, C_\text{Log}) = \sum_{i=1}^{29}\frac{1}{\sigma_i^2}\left(\frac{dN^i_\mathrm{obs}}{dT_{rec}} - \frac{dN^i_\mathrm{th}}{dT_{rec}} \right)^2 \, ,
\label{eq:chi}
\end{equation}
where the index $i$ runs over the $i$th bin of the observed XENON1T signal denoted by $dN^i_\mathrm{obs}/dT_{rec}$ with statistical uncertainty  $\sigma_i$, taken from Ref.~\cite{Aprile:2020tmw}. Here, $dN^i_\mathrm{th}/dT_{rec}(g_s, M_s, C_\text{Log})$ represents the expected signal due to hidden photon absorption as described in Eq.(\ref{eq:expected-signal}) including also the  background $B_0$.

Assuming effective dark photon parameters we perform a sensitivity analysis by varying simultaneously the mass and gauge coupling. The sensitivity contours in the parameter space $(m_X,~ g_{X,\text{eff}})$, are shown  in Fig.~\ref{fig:contour-eff} at $1\sigma$ and 95\% C.L. Also shown are the individual $\Delta \chi^2(m_X)$ and $\Delta \chi^2(g_{X,\text{eff}})$ functions, marginalized in each case over the undisplayed parameter. The best fit value in this case corresponds to $\chi^2_\text{min}=35.2$ for $g_{X,\text{eff}}= 9.3 \cdot 10^{-16}$  and $m_X=2.9$~keV. The XENON1T collaboration has also examined the possibility of an additional source of background coming from $\mathrm{^3H}$ decay. We are thus motivated to explore the impact of this new background  by taking into account a total $B_0 +\mathrm{^3H}$ background in the definition of Eq.(\ref{eq:chi}). As expected, the results imply a reduction of the significance with the best fit value being $\chi^2_\text{min}=38.14$ for $g_{X,\text{eff}}= 5.4 \cdot 10^{-16}$  and $m_X=2.6$~keV (the corresponding 1$\sigma$ allowed region is presented in Fig.~\ref{fig:contour-eff}). To compare the present results with existing astrophysical constraints, superimposed are limits coming from the cooling of horizontal branch stars (HB) and red giants (RGB) summarized in Ref.~\cite{Alonso-Alvarez:2020cdv} (see~\cite{Ayala:2014pea,Giannotti:2015kwo} for details).

Prompted by the stringent constraints derived in the context of the simplified effective scenario, i.e.  mainly due to the narrow range of allowed masses $m_X$,  we are now intended to probe the low scale string model parameters in the light of the recent XENON1T data. For simplicity, we reduce one degree of freedom by fixing $g_s$ to 0.02 or 0.2.  We note  that the latter choices ensure that $M_s>$10~TeV~\cite{Anchordoqui:2020tlp} and Eq.(\ref{eq:mx1}). For the two benchmark scenarios, in Fig.~\ref{fig:contour-string} we present the corresponding allowed regions at $1\sigma$ and 95\% C.L.  in the plane $(M_s,~ C_\text{Log})$ as well as the $\Delta \chi^2(M_s)$ and $\Delta \chi^2(C_\text{Log})$ functions.  Given the above considerations, we find two distinct regions with best fit values ($M_s= 991$~TeV, $C_\text{Log} =5.5 \cdot 10^{-10}$) for the case for $g_s=0.02$ and  ($M_s= 1770$~TeV, $C_\text{Log} =9.8 \cdot 10^{-10}$) for the case for $g_s=0.2$, both sharing a $\chi^2_\text{min} = 35.2$. For completeness, in the same figure also shown is the $1\sigma$ contour when the tritium background is taken into account.
Before closing our discussion, since our adopted scenario depends on three parameters $(g_s, M_s, C_\text{Log})$, we find it useful to perform a more generic sensitivity analysis, leaving free the full set of relevant parameters. Performing a scan in the region $g_s=(10^{-6},~ 10^{-1})$, maintaining this way the perturbativity of the theory, we marginalize over $g_s$ by evaluating $\chi^2(M_s, C_\text{Log}) = \text{min}_{g_s} \, \, \chi^2(g_s, M_s, C_\text{Log})$. The latter analysis led to the same best fit as in the previous cases, i.e. $\chi^2_\text{min}= 35.2$  which occurred at  ($M_s=153$~TeV, $C_\text{Log}=8.5 \cdot 10^{-11}$). We finally conclude that at 95\% C.L. the preferred values of $M_s$ lie in the range 70--2000~TeV while $C_\text{Log}$ falls in the $10^{-11}$--$10^{-9}$ range, for the various values of $g_{s}$.

\section{Conclusions}
\label{conclusion}
In this brief note we have focused on the hidden photon interpretation of the XENON1T excess, which seems to be very plausible for a number of reasons, not least of which is that a hidden photon dark matter relic with the mass around $2.8$ keV and a kinetic mixing of about $10^{-15}$ can not only explain the XENON1T excess but can also explain cooling excesses in HB stars.
Also the coupling of such a hidden photon to neutrinos is naturally highly suppressed by the $Z$ boson mass, making the hidden photon cosmologically stable. We have discussed how the very small masses and couplings can arise from  type-IIB low scale string theory.
We have also provided additional string theory motivation for weakly coupled gauge boson and hidden photons in D-brane models.

We have interpreted the XENON1T excess in terms of a minimal low scale type-IIB string theory parameter space and presented some benchmark points which describe well the data. We then provided a fit to the XENON1T signal, providing sensitivity contours in the plane of $m_X$ and $g_{X,\text{eff}}$, including the possibility of an additional source of background coming from $\mathrm{^3H}$ decay. To compare the present results with existing astrophysical constraints, we also included the limits coming from the cooling of horizontal branch stars (HB) and red giants (RGB). Finally we expressed the sensitivity plot of the XENON1T excess in the parameter space of $M_s$, $g_s$ and $C_\text{Log}$. The results show that the string scale must lie below about $2\times{10^{3}}$ TeV, providing some indication that string theory could be discovered at colliders in the not too distant future, if the XENON1T excess is due to a hidden photon resulting from string theory.

\section*{Acknowledgments}

 The work of D.K.P. is co-financed by Greece and the European Union (European Social Fund- ESF) through
the Operational Programme <<Human Resources Development, Education and
Lifelong Learning>> in the context of the project ``Reinforcement of
Postdoctoral Researchers - 2nd Cycle" (MIS-5033021), implemented by
the State Scholarships Foundation (IKY). G.K.L. would like to thank Ignatios Antoniadis for  clarifications regarding~\cite{Anchordoqui:2020tlp}. S.F.K. acknowledges the STFC Consolidated Grant No. ST/L000296/1 and the European Union's Horizon 2020 Research and Innovation programme under Marie Sk\l {}odowska-Curie grant agreements Elusives ITN No.\ 674896 and InvisiblesPlus RISE No.\ 690575.

\appendix

 
 \newpage

\section{List of solutions} 
\label{app}

Table~\ref{tab:solutions1} displays all the solutions with $c_{2}=c_{3}=0$ for the brane configuration described by the left panel in Fig.~\ref{fig1:Dbranemodels}. There are 32 solutions in this class. 

\begin{table}[h]
\centering\resizebox{0.66\textwidth}{!}{%
\begin{tabular}{ccccccc|ccccc}\hline\hline
$\eps_1$ & $\eps_2$ & $\eps_3$ & $\eps_4$ & $\eps_5$ &$\eps_6$ & $\eps_7$ & $k_{3}$ & $k_{2}$ & $c_{1}$ & $c_{2}$ & $c_{3}$
 \\\hline
  -1 & -1 & -1 & -1 & -1 & -1 & -1 & 2/3 & 1/2 & -1 & 0 & 0 \\
 -1 & -1 & -1 & -1 & -1 & -1 & 1 & 2/3 & 1/2 & -1 & 0 & 0 \\
 -1 & -1 & -1 & -1 & 1 & -1 & -1 & 2/3 & 1/2 & -1 & 0 & 0 \\
 -1 & -1 & -1 & -1 & 1 & -1 & 1 & 2/3 & 1/2 & -1 & 0 & 0 \\
 -1 & -1 & 1 & -1 & -1 & 1 & -1 & 2/3 & 1/2 & 1 & 0 & 0 \\
 -1 & -1 & 1 & -1 & -1 & 1 & 1 & 2/3 & 1/2 & 1 & 0 & 0 \\
 -1 & -1 & 1 & -1 & 1 & 1 & -1 & 2/3 & 1/2 & 1 & 0 & 0 \\
 -1 & -1 & 1 & -1 & 1 & 1 & 1 & 2/3 & 1/2 & 1 & 0 & 0 \\
 -1 & 1 & -1 & -1 & -1 & -1 & -1 & 2/3 & 1/2 & -1 & 0 & 0 \\
 -1 & 1 & -1 & -1 & -1 & -1 & 1 & 2/3 & 1/2 & -1 & 0 & 0 \\
 -1 & 1 & -1 & -1 & 1 & -1 & -1 & 2/3 & 1/2 & -1 & 0 & 0 \\
 -1 & 1 & -1 & -1 & 1 & -1 & 1 & 2/3 & 1/2 & -1 & 0 & 0 \\
 -1 & 1 & 1 & -1 & -1 & 1 & -1 & 2/3 & 1/2 & 1 & 0 & 0 \\
 -1 & 1 & 1 & -1 & -1 & 1 & 1 & 2/3 & 1/2 & 1 & 0 & 0 \\
 -1 & 1 & 1 & -1 & 1 & 1 & -1 & 2/3 & 1/2 & 1 & 0 & 0 \\
 -1 & 1 & 1 & -1 & 1 & 1 & 1 & 2/3 & 1/2 & 1 & 0 & 0 \\
 1 & -1 & -1 & 1 & -1 & -1 & -1 & 2/3 & -1/2 & -1 & 0 & 0 \\
 1 & -1 & -1 & 1 & -1 & -1 & 1 &  2/3 & -1/2  & -1 & 0 & 0 \\
 1 & -1 & -1 & 1 & 1 & -1 & -1 &  2/3 & -1/2  & -1 & 0 & 0 \\
 1 & -1 & -1 & 1 & 1 & -1 & 1 &  2/3 & -1/2  & -1 & 0 & 0 \\
 1 & -1 & 1 & 1 & -1 & 1 & -1 &  2/3 & -1/2  & 1 & 0 & 0 \\
 1 & -1 & 1 & 1 & -1 & 1 & 1 &  2/3 & -1/2  & 1 & 0 & 0 \\
 1 & -1 & 1 & 1 & 1 & 1 & -1 &  2/3 & -1/2  & 1 & 0 & 0 \\
 1 & -1 & 1 & 1 & 1 & 1 & 1 &  2/3 & -1/2  & 1 & 0 & 0 \\
 1 & 1 & -1 & 1 & -1 & -1 & -1 &  2/3 & -1/2  & -1 & 0 & 0 \\
 1 & 1 & -1 & 1 & -1 & -1 & 1 &  2/3 & -1/2  & -1 & 0 & 0 \\
 1 & 1 & -1 & 1 & 1 & -1 & -1 &  2/3 & -1/2  & -1 & 0 & 0 \\
 1 & 1 & -1 & 1 & 1 & -1 & 1 &  2/3 & -1/2  & -1 & 0 & 0 \\
 1 & 1 & 1 & 1 & -1 & 1 & -1 &  2/3 & -1/2  & 1 & 0 & 0 \\
 1 & 1 & 1 & 1 & -1 & 1 & 1 &  2/3 & -1/2  & 1 & 0 & 0 \\
 1 & 1 & 1 & 1 & 1 & 1 & -1 &  2/3 & -1/2  & 1 & 0 & 0 \\
 1 & 1 & 1 & 1 & 1 & 1 & 1 &  2/3 & -1/2  & 1 & 0 & 0 \\\hline\hline
 \end{tabular}%
}
\caption{\small{Analytic solutions with $c_{2}=c_{3}=0$. }}
\label{tab:solutions1}
\end{table}

\bibliography{bibliography} 

\end{document}